\documentclass[pra,aps,amsmath,amssymb,preprint]{revtex4-1}
\usepackage{dcolumn}
\usepackage{bm}
\usepackage{color}
\usepackage{rotating}
\usepackage{graphicx}
\usepackage{latexsym}
\usepackage{amssymb}
\usepackage{amsfonts}
\usepackage{soul}
\usepackage{color}
\usepackage{amsmath}
\usepackage[T2A]{fontenc}
\usepackage[utf8]{inputenc}

\begin{document}

\title{Room-temperature anisotropic in-plane spin dynamics in graphene induced by PdSe$_2$ proximity}

\author{Juan F. Sierra$^{1,\S}$}
\email{juan.sierra@icn2.cat}
\author{Josef Sv\v{e}tl\'{\i}k$^{1,2,\S}$}
\email{josef.svetlik@icn2.cat}

\author{Williams Savero Torres$^{1}$}
\author{Lorenzo Camosi$^{1}$}
\author{Franz Herling$^{1}$}
\author{Thomas Guillet$^{1}$}
\author{Kai Xu$^{3}$}
\author{Juan Sebasti\'{a}n Reparaz$^{3}$}
\author{Vera Marinova$^{4}$}
\author{Dimitre Dimitrov$^{4,5}$}
\author{Sergio O. Valenzuela$^{1,6}$}
\email{SOV@icrea.cat}
\thanks{\\$^\S$These authors contributed equally to this work}

\renewcommand{\baselinestretch}{1.2}\normalsize
\affiliation{$^1${\small Catalan Institute of Nanoscience and Nanotechnology (ICN2), CSIC and The Barcelona Institute of Science and Technology (BIST),
 Campus UAB, Bellaterra, 08193 Barcelona, Spain}}
\affiliation{$^2$ {\small Universitat Aut\`{o}noma de Barcelona, Bellaterra, 08193 Barcelona, Spain}}
\affiliation{$^3$ {\small Institut de Ci\`{e}ncia de Materials de Barcelona, \\
ICMAB-CSIC, Campus UAB, 08193 Bellaterra, Spain}}
\affiliation{$^4$ {\small Institute of Optical Materials and Technologies,\\
 Bulgarian Academy of Science, 1113 Sofia, Bulgaria}}
\affiliation{$^5$ {\small Institute of Solid State Physics, Bulgarian Academy of Sciences, 1784 Sofia, Bulgaria}}
\affiliation{$^6$ {\small Instituci\'{o} Catalana de Recerca i Estudis Avan\c{c}ats (ICREA), 08010 Barcelona, Spain}}

\date{\today}

\begin{abstract}
\vspace{2mm}
\renewcommand{\baselinestretch}{1.32}\normalsize
\textbf{Van der Waals heterostructures provide a versatile platform for tailoring electrical, magnetic, optical, and spin transport properties via proximity effects. Hexagonal transition metal dichalcogenides induce valley-Zeeman spin-orbit coupling (SOC) in graphene, creating spin lifetime anisotropy between in-plane and out-of-plane spin orientations. However, in-plane spin lifetimes remain isotropic due to the inherent heterostructure's threefold symmetry. Here, we demonstrate that pentagonal PdSe$_2$, with its unique in-plane anisotropy, induces anisotropic gate-tunable SOC in graphene. This enables a 10-fold modulation of spin lifetimes at room temperature, depending on the in-plane spin orientation. Moreover, the directional dependence of the spin lifetimes, along the three spatial directions, reveals a persistent in-plane spin texture component that governs the spin dynamics. These findings advance our understanding of spin physics in van der Waals heterostructures and pave the way for designing topological phases in graphene-based heterostructures in the strong SOC regime.}

\end{abstract}

\maketitle

Spin-orbit coupling (SOC) is pivotal in modern condensed matter physics, enabling the realization of unique topological phases \cite{RMP2010a,RMP2010b}, charge-to-spin interconversion phenomena \cite{SOV2015}, spin-orbit torques \cite{manchon2019} and spin qubit manipulation \cite{nadj-perge2010,brooks2020}. Its interplay with superconductivity, potentially resulting in Majorana zero modes and triplet pairing, further highlights its broad significance \cite{Majoranareview,amundsen2024}. Therefore, mastering the design and control of SOC is crucial for advancing our understanding of quantum phenomena and harnessing them for novel technological applications. Crucially, this can be achieved via the van der Waals (vdW) proximity effect without affecting other electronic properties \cite{zutic2020,JFS2021}.

Proximity effects, extensively investigated in superconductivity, gain prominence in vdW materials owing to their atomic-scale thickness. This attribute enables the exploration of proximity phenomena involving short-range interactions, like magnetism and SOC, within vdW heterostructures. Magnetic proximity effects have been studied in various vdW materials. For instance, graphene and 2H transition metal dichalcogenides (TMDCs) exhibit magnetism when interfaced with vdW magnets like CrSBr, CI$_3$ or CrBr$_3$ \cite{ghiasiMPE,zhongMPE,lyons2020}. Proximity-induced SOC has been proposed to stabilise perpendicular magnetic anisotropy in vdW ferromagnets \cite{wang2020}, enhance bilayer graphene's superconducting critical temperature \cite{zhangSC2023}, and improve THz detector sensitivity \cite{KS2024}. Hybrid graphene-2H TMDCs systems are particularly promising for spin and opto-spin phenomena \cite{gmitra2015,gmitra2016}, where proximity-SOC has been demonstrated through spin relaxation anisotropy \cite{ghiasi2017,LAB2018} and charge-to-spin interconversion (CSI) experiments \cite{safeer2019,ghiasi2019,LAB2020,RG2021}. Giant spin relaxation anisotropy persists up to room temperature \cite{ghiasi2017,LAB2018,herling2020}, attributed to valley-Zeeman SOC \cite{cummings2017,offidani2018}. While Rashba-type SOC has been observed, in-plane spin relaxation remains isotropic due to the heterostructure $\mathcal{C}_3$ symmetry. Unconventional CSI has been reported in graphene combined with low-symmetry semimetals MoTe$_2$ \cite{safeer2019b,hoque2021} and WTe$_2$ \cite{zhao2020,LC2022}, although bulk contributions from the semimetals complicate isolating graphene-specific effects.

In-plane spin lifetime anisotropy has not been conclusively demonstrated, though low-temperature studies of anisotropic black phosphorus suggest potential for directionally tunable spin transport \cite{cording2024}. Graphene, with its superior spin transport properties, could enhance these effects when combined with proximity-SOC, enabling their observation at higher temperatures with improved efficiency.  Heterostructures integrating materials with disparate crystalline symmetries are particularly relevant in this context, but remain unexplored \cite{newfabian}. Here, we investigate spin transport in graphene-PdSe$_2$ heterostructures. PdSe$_2$, a semiconducting TMDC, features a band gap of about 1.3 eV in monolayer form that decreases with thickness \cite{sun2015,oyedele2017}. Its unit cell is orthorhombic, comprising buckled monolayers of irregular pentagons aligned along the $a$-axis (Fig. 1a).

The valley-Zeeman SOC imprinted into graphene by TMDCs is attributed to interlayer tunnelling \cite{david2019}. In this band-to-band tunnelling model \cite{david2019}, electrons in graphene can tunnel to any of the three atomic layers of PdSe$_2$, but the probability of reaching the first Se layer is exponentially higher. Thus, the buckled structure of PdSe$_2$, with its anisotropic Se atom arrangement(Fig. 1a), could generate novel spin textures with strong in-plane anisotropy. To test this hypothesis, we characterised spin lifetimes in monolayer graphene-PdSe$_2$ along three orthogonal axes using the non-local spin device design illustrated in Fig. 1b. The experiments systematically explore in-plane and out-of-plane spin precession with varying magnetic field $\mathbf{B}$ orientations \cite{BR2016,BR2017,LAB2019} relative to the $\hat{x}$, $\hat{y}$ and $\hat{z}$ axes, where $\hat{z}$ is perpendicular to the graphene plane and $\hat{y}$ aligns with the ferromagnetic electrodes, F1-F4. A PdSe$_2$ single crystal is placed off-centre between F1 and F2, which serve as either spin injector or spin detector. PdSe$_2$ introduces proximity SOC in graphene, modifying spin dynamics and relaxation. The spin lifetimes $\tau_{\hat{x}', \hat{y}',\hat{z}'}^{\mathrm{s}}$ along $\hat{x}'$, $\hat{y}'$ and $\hat{z}'=\hat{z}$, correlate with the induced SOC and spin textures. Here, $\hat{x}'$ ($\hat{y}'$) corresponds to the in-plane direction with the longest (shortest) spin lifetime. 
Due to the heterostructure's low symmetry, the spin textures may not align with the graphene or PdSe$_2$ crystalline axes. The angle $\theta$ characterises the $\hat{x}'$ orientation relative to the PdSe$_2$ crystallographic axis $\hat{a}$, determined by Raman spectroscopy.

In our devices, the distance between F1 and F2 is approximately 10 $\mu$m. This distance is sufficient to ensure complete dephasing of the spin accumulation perpendicular to $\mathbf{B}$ at moderately low $B$ for spins diffusing between F1 and F2 \cite{BR2016}. Two additional ferromagnetic electrodes, F3 and F4, along with F1 and F2, form reference devices for characterising the spin dynamics in pristine graphene. A back-gate voltage $V_\mathrm{g}$ applied to the Si/SiO$_{2}$ substrate tunes the carrier density in the graphene-PdSe$_2$ heterostructure (Supplementary Fig. 1). An optical image of a typical device (Device 1) is shown in Fig. 1c (see Methods for fabrication details). The PdSe$_2$ crystal is $1.4$ $\mu$m wide and $20$-nm thick. Figure 1d illustrates representative Raman spectra of this crystal under parallel polarization for rotation angles between $0^\circ$ and $360^\circ$ relative to $\hat{x}$. The A$_\mathrm{g}$ and B$_{\mathrm{1g}}$ modes are observed, with the A$_\mathrm{g}$ peaks displaying intensity modulation with a $180^\circ$ period, enabling the identification of the crystalline axes \cite{yu2020}. The angular dependence of the A$_\mathrm{g}^2$ mode intensity in Fig. 1e shows that the $a$ axis aligns with the long crystal side (see Fig. 1c), with a 20$^\circ$ counterclockwise tilt from $\hat{y}$.

Figure 2 illustrates the dramatic impact of PdSe$_2$ on graphene spin dynamics. The nonlocal resistance, $R_{\mathrm{nl}}=V_{\mathrm{nl}}/I$, is determined from the voltage $V_{\mathrm{nl}}$ at the detector generated by a current $I$ at the injector (Fig. 1b). To investigate the in-plane spin-lifetime anisotropy, $\mathbf{B}$ is sequentially swept along the $\hat{z}$, $\hat{y}$ and $\hat{x}$ directions while monitoring $R_{\mathrm{nl}}$. Figures 2a-c show $R_{\mathrm{nl}}$ for pristine graphene, obtained using the reference graphene device defined by F1 and F3, while Figs. 2d-g present the corresponding measurements for the graphene-PdSe$_2$ device defined by F1 and F2. All measurements were performed at room temperature.

The magnetisations of the ferromagnetic electrodes align along their length, in the $\hat{y}$ direction, owing to shape anisotropy. Therefore, $B_{z}$ induces spin precession exclusively in the $x-y$ plane (see Fig. 2g). Since spin relaxation in graphene on SiO$_2$ is isotropic \cite{BR2016,BR2017,LAB2019}, the spin dynamics in the reference device does not depend on the spin orientation, and the spin precession lineshape is symmetric about $B_z=0$ (Fig. 2a). In contrast, the spin precession profile in graphene-PdSe$_2$ is strongly asymmetric (Fig. 2d), with $R_{\mathrm{nl}}$ at the extrema (at $B_{z}\approx \pm20$ mT) differing by a factor $\sim 3.5$. This suggests that the spin relaxation in the graphene channel beneath PdSe$_2$ depends on the direction of precession, indicating strong in-plane spin relaxation anisotropy. It also shows that neither $\hat{x}'$ nor $\hat{y}'$ align with $\hat{y}$ (the orientation of injected spins), otherwise the precession profile would be symmetric about $B_z=0$, albeit with a different lineshape than in Fig. 2a.

Further evidence of in-plane spin relaxation anisotropy is displayed in Figs. 2b and 2e. When $\mathbf{B}$ is aligned to $\hat{y}$ (Fig. 2h), the injected spins, which are parallel to $\mathbf{B}$, do not precess in the isotropic graphene of the reference device, making $R_{\mathrm{nl}}$ independent of $B_y$ (Fig. 2b). In contrast, $R_{\mathrm{nl}}$ becomes field-dependent in graphene-PdSe$_2$ (Fig. 2e). This behaviour arises because the injected spins along $\hat{y}$ are misaligned with $\hat{x}'$ and $\hat{y}'$. Since $\tau_{x'}^{\mathrm{s}}>\tau_{y'}^{\mathrm{s}}$, spins in graphene-PdSe$_2$ relax and shift towards $\hat{x}'$, losing the alignment with $\hat{y}$ even for $B=0$ (Fig. 2h). Thus, when $\mathbf{B}$ is applied along $\hat{y}$, the spins are no longer parallel to $\mathbf{B}$ and start precessing around $\hat{y}$, acquiring an out-of-plane component. As a result, the signal becomes dependent on $B_y$, influenced by $\tau_{x'}^{\mathrm{s}}$, $\tau_{y'}^{\mathrm{s}}$, and $\tau_{z'}^{\mathrm{s}}$.

In-plane anisotropy is also revealed by applying $\mathbf{B}$ along $\hat{x}$ (Figs. 2c and 2f). For sufficiently large $B_x$, the F1-F4 magnetisations rotate and align parallel to $\hat{x}$ at the saturation magnetic field $B_x^\mathrm{sat}$ (see Fig. 2i). Since $B$ lies within the graphene plane, magnetoresistance effects have no influence \cite{BR2016,BR2017}. In isotropic graphene, $R_{\mathrm{nl}}(B=0)\approx R_{\mathrm{nl}}(B_x>B_x^\mathrm{sat})$ (Fig. 2c). In graphene-PdSe$_2$, however, due to in-plane anisotropy, $R_{\mathrm{nl}}(B=0)$ and $R_{\mathrm{nl}}(B_x>B_x^\mathrm{sat})$ typically differ (Fig. 2f). The precession profile depends on spin lifetimes in all orthogonal directions. As with Fig. 2e, the misalignment between $\hat{x}$ and both $\hat{x}'$ and $\hat{y}'$ leads to a signal dependent on $B_x$, even for $B_x>B_x^\mathrm{sat}$.

To quantify the spin lifetime anisotropy in graphene-PdSe$_2$, we estimate $\tau_{x', y',z}^{\mathrm{s}}$ by modelling the experimental results using the solution of the Bloch diffusion equations. The spin lifetime $\tau^\mathrm{s}$ and diffusion constant $D^\mathrm{s}$ in pristine graphene are determined from spin-precession measurements under an applied $B_z$ in the reference device (Fig. 2c, see also Supplementary Fig. 3). These parameters are then used to extract $\theta$ and $\tau_{x', y'}^\mathrm{s}$ from the corresponding measurements in graphene-PdSe$_2$ (Fig. 2d). An initial estimate of $\tau_{z}^\mathrm{s}$ is obtained using the data in Fig. 2e, with $\tau_{x', y'}^\mathrm{s}$ and $\theta$ fixed. A more precise estimate of $\tau_{z}^\mathrm{s}$ can be derived by forcing the spins to precess out of the plane, for instance, by aligning $\mathbf{B}$ with $\hat{x}$ (Fig. 2g). However, as discussed in Ref. \cite{BR2017}, modelling the spin precession profile in this configuration is challenging due to the substantial and non-uniform rotation of the F1/F2 magnetisations with $B_x$, even under moderate $B_x$ conditions. A more reliable method to determine $\tau_{x', y',z}^{\mathrm{s}}$ involves applying an oblique $\mathbf{B}$, characterised by an angle $\beta$, as illustrated in Fig. 3a \cite{BR2016,BR2017}. This magnetic field induces spin precession tracing out a cone. By varying $\beta$ and the magnitude of $\mathbf{B}$, the spins reaching the graphene-PdSe$_2$ region change their orientation from 0 to 180$^\circ$ relative to the graphene plane. This allows us to systematically scan the spin dynamics across all possible spin orientations and adjust the values of $\tau_{x', y',z}^{\mathrm{s}}$.

Figures 3a and 3b show $R_{\mathrm{nl}}$ for representative $\beta$, using F1(F2) and F2(F1) as injector (detector), respectively. The interchange of the F1/F2 roles alters the spin orientation in the graphene-PdSe$_2$ region for a given $\beta$ and $B$, due to the non-equidistant positioning of PdSe$_2$ relative to F1 and F2. This provides an additional knob to validate the diffusive model. The two configurations exhibit complete dephasing at $B=B_\mathrm{d}\approx 0.1$ T, but differences in precession profiles are evident. Specifically, when F1 is the injector (Fig. 3a) the $R_{\mathrm{nl}}$ extremum at $B\approx 20$ mT is less pronounced compared to when F2 is the injector (Fig. 3b). The difference is most pronounced at $\beta = 90^\circ$. As $\beta$ decreases, $R_{\mathrm{nl}}$ shows an anomalous decrease even for $B>B_\mathrm{d}$, in contrast to pristine graphene, where $R_{\mathrm{nl}}$ approaches a constant value and then slowly increases due to the out-of-plane tilting of the F1/F2 magnetizations at high $B$ \cite{BR2016,BR2017}.

The solid lines in Figs. 2 and 3 represent the solution of Bloch diffusion equations for $\tau_{x'}^{\mathrm{s}}= 260$ ps, $\tau_{y'}^{\mathrm{s}}=21$ ps, $\tau_{z}^{\mathrm{s}}=18.5$ ps and $\theta=71^\circ$. In Fig. 3, only $\beta$ is varied for each curve. The agreement between the model and experimental data is remarkable, capturing all trends, which is especially noteworthy given the wide range of measurements and experimental conditions. To fit the results in Fig. 2f, the Stoner-Wohlfarth approximation is adopted with an in-plane saturation magnetic field $B_x^\mathrm{sat}=0.18$ T. The diffusion model precisely accounts for subtle differences observed for $B_x<0$ and $B_x>0$, attributed to the rotation of the ferromagnets' magnetisation.

The spin lifetimes $\tau_{x', y',z}^{\mathrm{s}}$ were also determined as a function of $V_\mathrm{g}$. Although tunable spin relaxation anisotropy has been predicted in 2H-TMDCs-graphene heterostructures \cite{cummings2017}, systematic experimental studies are lacking in the literature \cite{ghiasi2017,LAB2018}. The results, summarized in Fig. 3c (see also Supplementary Fig. 4), reveal that the spin dynamics remains anisotropic, over the measured $V_\mathrm{g}$ range. The in-plane anisotropy ratio, $\zeta_{xy}\equiv\frac{\tau_{x'}}{\tau_{y'}}$, decreases from $\zeta_{xy}\approx12.4$ at $V_\mathrm{g}=-30$ V to $\zeta_{xy}\approx2.5$ at $V_\mathrm{g}=25$ V. This decrease is driven by changes in $\tau_{x'}^{\mathrm{s}}$, while $\tau_{y'}^{\mathrm{s}}$ and $\tau_{z}^{\mathrm{s}}$ remain roughly constant, with $\zeta_{zy}\equiv\frac{\tau_{z}^{\mathrm{s}}}{\tau_{y'}^{\mathrm{s}}} \approx 1.1-1.3$.

The measurements and analysis in Figs. 2 and 3 conclusively establish that the anisotropic spin relaxation is an interfacial effect induced by the proximity of PdSe$_2$ to graphene, rather than from spin absorption. This is supported by three key observations: \textit{i}) the significant misalignment between the primary spin relaxation axes, $\hat{x}'$ and $\hat{y}'$, and the PdSe$_2$ crystalline axes ($\theta = 71^\circ \neq 0, 90^\circ$, see Fig. 1e), \textit{ii}) the excellent agreement between our experimental data and a spin diffusion model that does not consider spin absorption, and \textit{iii}) the lack of correlation between the gate dependence of the spin lifetimes and anisotropy ratios and the gate-dependent transport properties of PdSe$_2$. While spin absorption in PdSe$_2$ could potentially affect spin lifetimes and anisotropy ratios, this scenario is unlikely. The 20-nm thick PdSe$_2$ layer in Device 1 has a small bandgap ($\sim$ 30 meV) and a very low conductivity $\sigma_\mathrm{PdSe_2}$ at room temperature \cite{sun2015}. PdSe$_2$ exhibits ambipolar transport, transitioning from hole to electron carriers across the $V_\mathrm{g}$ range in Fig. 3c, with minimum conductivity at $V_\mathrm{g}\approx-12$ V (Supplementary Fig. 5c). However, no correlation is observed between the gate-invariant $\tau_{y'}^{\mathrm{s}}$ and $\tau_{z}^{\mathrm{s}}$, or the monotonically decreasing $\tau_{x'}^{\mathrm{s}}$ and $\zeta_{xy}$, and the transport properties of PdSe$_2$, which would be expected if spin absorption were a contributing factor.

To validate these conclusions, we characterized spin transport at 77 K in a graphene-PdSe$_2$ device with a thinner 6-nm PdSe$_2$ layer (Device 2) and an estimated band gap of over $500$ meV), where charge transport is expected to vanish \cite{oyedele2017}. An optical image of the device is shown in Fig. 4a (inset). The angular dependence of the A$_\mathrm{g}^2$ Raman mode intensity (Fig. 4b) identifies the $a$ axis of PdSe$_2$. Measurements of the current $I_{\mathrm{ds}}$ versus the driving voltage $V_{\mathrm{ds}}$ applied between graphene and PdSe$_2$ (Fig. 4a) reveal that $I_{\mathrm{ds}}$ is undetectable across -50 V $<V_\mathrm{g}<$ 50 V at negative and low $V_{\mathrm{ds}}$, sharply increasing only when $V_{\mathrm{ds}}>0.2$ V with positive $V_{\mathrm{g}}$. This behaviour confirms that the PdSe$_2$ band gap suppresses charge transfer between graphene and PdSe$_2$ \cite{yan2016,dankert2017}.

Figures 4c-4f detail the spin dynamics of Device 2. Figure 4c shows $R_{\mathrm{nl}}$ versus $B_z$. The spin precession profile, while symmetric, notably deviates from the typical profile observed in pristine graphene (cf. Fig. 2a), with $|R_{\mathrm{nl}}|$ decreasing to zero without a sign change as $|B_z|$ increases. This suggests $\hat{x}'$ is nearly aligned with $\hat{y}$, the orientation of the injected spins. This alignment is corroborated by the weak $R_{\mathrm{nl}}$ variation with $B_y$ (Fig. 4d) and sharp suppression of $R_{\mathrm{nl}}$ under $B_x$, which rotates the F1/F2 magnetizations towards $\hat{x}$ (Fig. 4e). For $B_x>B_x^\mathrm{sat}$, $R_{\mathrm{nl}}$ decreases to zero, demonstrating a more pronounced reduction than in Device 1 (cf. Fig. 3f). This stronger suppression is attributed to the larger PdSe$_2$ width, spanning 2.8 $\mu$m of the graphene channel compared to 1.4 $\mu$m in Device 1, which increases the contrast in the exponential decay for spins along $\hat{x}'$ and $\hat{y}'$, characterised by distinct relaxation lengths. Spin lifetimes and $\theta$ were extracted from oblique spin precession (Fig. 4f) and polarized Raman measurements (Fig. 4b), yielding $\tau_{x'}^{\mathrm{s}}= 160$ ps, $\tau_{y'}^{\mathrm{s}}=20$ ps, $\tau_{z}^{\mathrm{s}}=5$ ps and $\theta=18^\circ$. These spin lifetimes and $\zeta_{xy}\approx 8$ are consistent with those in Device 1. Solid lines in Figs. 4c-4f, derived from the diffusion model using these parameters, show excellent agreement with the experimental data.

The results from Device 2 confirm the reproducibility of our findings. The comparable anisotropy between Devices 1 and 2 further demonstrates that spin absorption plays a negligible role. The substantial misalignment between the PdSe$_2$ $a$ axis and $\hat{x}'$ in both devices ($71^\circ$ in Device 1 and $18^\circ$ in Device 2) indicates no direct correlation between spin relaxation anisotropy and the PdSe$_2$ crystalline axes and underscores the crucial role of proximity effects in driving the observed anisotropy. The spin-lifetime hierarchy $\tau_{x'}^{\mathrm{s}}\gg\tau_{y'}^{\mathrm{s}}\gtrsim\tau_z^{\mathrm{s}}$ suggests the presence of a momentum-independent (persistent) spin texture along $\tau_{x'}^{\mathrm{s}}$, which dominates spin relaxation. Such textures are predicted in low-symmetry structures like monolayers of topological insulators 1T$_\mathrm{d}$-MoTe$_2$ and 1T$_\mathrm{d}$-WTe$_2$ \cite{MV2021}. Significant in-plane spin relaxation anisotropy is expected in the conduction band of 1T$_\mathrm{d}$-MoTe$_2$ monolayers near the band edge \cite{MV2021}, although this remains experimentally unverified. A novel SOC term predicted in graphene-SnTe, linked to the crystal gradient potential of low-symmetry SnTe \cite{newfabian}, would lead to a dominant in-plane spin texture in proximitized graphene, polarized nearly perpendicular to this gradient. Calculated spin lifetimes in graphene-SnTe reveal a large anisotropy below the valence band maximum, closely mirroring the hierarchy observed in our study.

Experimentally, in-plane spin transport anisotropy has been reported in ultrathin black phosphorus at low temperatures (1.5 K), showing modest $\zeta_{xy}\approx 1-1.8$ and a larger $\zeta_{zy}\approx 6$ \cite{cording2024}. However, the spin lifetime hierarchy $\tau_{z}^{\mathrm{s}}\gg\tau_{x'}^{\mathrm{s}}\gtrsim\tau_{y'}^{\mathrm{s}}$ differs from that of graphene-PdSe$_2$, resembling instead that of graphene proximitized by 2H-TMDCs. In addition, the anisotropy in black phosphorous remains unexplained, as Elliot-Yafet-dominated spin relaxation should yield  $\tau_z^{\mathrm{s}}<\tau_{x',y'}^{\mathrm{s}}$ \cite{kurpas2018}.

Spin dynamics in proximitized graphene follows the Dyakonov-Perel mechanism \cite{cummings2017,ghiasi2017,LAB2018}. The spin relaxation rates are given by $(\tau_{i}^{s})^{-1} = [\langle\Omega^2_{j}\rangle_p+\langle\Omega^2_{k}\rangle_p] \tau_p + [\langle\Omega^2_{j}\rangle_{\mathrm{iv}}+\langle\Omega^2_{k}\rangle_{\mathrm{iv}}] \tau_{\mathrm{iv}}$ with $\{i, j, k\}\subseteq \{x',y',z\}$, where $\tau_p$ and $\tau_{\mathrm{iv}}$ represent respectively the momentum and intervalley scattering times, and $\langle\Omega^2_{i}\rangle_{p,\mathrm{iv}}$ are the momentum-dependent (index $p$) and momentum-independent (index iv) effective spin-orbit field components. In graphene-TMDC structures preserving $\mathcal{C}_3$ symmetry, $\langle\Omega^2_{i}\rangle_{p}$ arise from the conventional Rashba and pseudospin inversion asymmetry SOC terms with $\langle\Omega^2_{z}\rangle_{p}=0$, while $\langle\Omega^2_{i}\rangle_{\mathrm{iv}}$ stems from the valley-Zeeman SOC with $\langle\Omega^2_{x'}\rangle_\mathrm{iv}=\langle\Omega^2_{y'}\rangle_\mathrm{iv}=0$ \cite{cummings2017}. Out-of-plane spins relax via Rashba SOC, while the in-plane spins are affected by both the Rashba and valley-Zeeman SOCs. The resulting spin lifetime anisotropy ratio is $\zeta_{zx(y)}\equiv\frac{\tau_{z}^{\mathrm{s}}}{\tau_{x(y)}^{\mathrm{s}}}= (\lambda_\mathrm{vZ}/\lambda_\mathrm{R})^2 \frac{\tau_\mathrm{iv}}{\tau_p}+1/2$, where $\lambda_\mathrm{R}$ and $\lambda_\mathrm{vZ}$ are the Rashba (R) and valley-Zeeman (vZ) SOC strengths \cite{cummings2017}. Since typically $\frac{\tau_\mathrm{iv}}{\tau_p}\gg1$ and $\lambda_\mathrm{R} \sim \lambda_\mathrm{vZ}$, $\zeta_{zx(y)}\gg1$.  In graphene-PdSe$_2$, the momentum-independent spin-orbit field aligns with $x'$, thus $\langle\Omega^2_{y'}\rangle_\mathrm{iv}=\langle\Omega^2_{z'}\rangle_\mathrm{iv}=0$. Spin relaxation along $x'$ is dominated by Rashba SOC, while along $y'$ and $z$, both the Rashba SOC and the persistent spin-orbit field contribute. This leads to $\zeta_{xy}=2(\lambda_\mathrm{pers}/\lambda_\mathrm{R})^2 \frac{\tau_\mathrm{iv}}{\tau_p}+1$ and $\zeta_{zy} \equiv \frac{\tau_{z}^{\mathrm{s}}}{\tau_{y}^{\mathrm{s}}}\lesssim 1$.

To estimate $\lambda_\mathrm{R}$ and $\lambda_\mathrm{pers}$, knowledge of the characteristic timescales $\tau_p$ and $\tau_\mathrm{iv}$ is required. Intervalley scattering time in graphene, determined through spin relaxation anisotropy experiments \cite{ghiasi2017,LAB2018} and weak localization measurements \cite{morpurgo2015,morpurgo2016,bouchiat2018}, consistently fall within the 5-20 ps range, regardless of the substrate or fabrication process. Assuming $\tau_\mathrm{iv}\approx10$ ps and $\tau_p\approx0.4$ ps (from Device 1 conductivity), the experimental $\zeta_{xy} \approx 12.5$ for graphene-PdSe$_2$ corresponds to $\lambda_\mathrm{pers}/\lambda_\mathrm{R}\approx0.5$. With $\lambda_\mathrm{R}$ conservatively estimated at 5 meV, based on graphene-2H-TMDC systems where $\lambda_\mathrm{R}$ is in the range of 5 and 30 meV \cite{LAB2020,morpurgo2016,monaco2021,wang2019}, we deduce $\lambda_\mathrm{pers} \approx 2.5$ meV.

The origin of $\theta$ remains an open question, requiring density functional theory studies in graphene-PdSe$_2$ heterostructures, similar to those performed for graphene-SnTe \cite{newfabian}. However, predicting $\theta$ as a function of twist angles is challenging. Even in simpler proximity systems, such as graphene coupled with 2H-TMDCs, the rotation angle of the Rashba SOC spin texture away from tangential is highly sensitive to the heterostructure's properties, with no discernible trend observed with varying twist angle \cite{naimer2021}. Nonetheless, the observed spin-lifetime anisotropy in graphene-PdSe$_2$ highlights the versatility of proximity effects in van der Waals heterostructures, providing valuable insights into strategies for tailoring spin textures through the interplay of materials with distinct crystal symmetries. The detection of persistent spin textures in graphene-PdSe$_2$, resembling those predicted for monolayer 1T$_d$-MoTe$_2$ and 1T$_d$-WTe$_2$, unveils promising avenues for advancing spin manipulation and exploring topologically non-trivial states driven by SOC. Furthermore, the tunability of spin relaxation anisotropy opens exciting prospects for the development of directionally controllable spin transport, which could have significant implications for spintronic applications.

\vspace{5mm}

\noindent \textbf{Acknowledgments}

\noindent We thank J.H. Garc\'{\i}a, A.W. Cummings and S. Roche for insightful discussions. This research was partially supported by MICIU/AEI/10.13039/501100011033 through Grants No. PID2022-143162OB-I00 and Severo Ochoa Programme CEX2021-001214-S, FEDER, and by the European Union's Horizon 2020 (H2020) FET-PROACTIVE project TOCHA under grant agreement 824140. ICN2 acknowledges support from PCI2019-103739 funded by MICIU/AEI/10.13039/501100011033 and V.M. and D.D. the support from the Bulgarian Science Fund under the grants DFNI KП-06-ДO 02/2 and DFNI KП-06-ДO 02/3, within the framework of the M-ERA program project 2D-SPIN-MEM “Functional 2D materials and heterostructures for hybrid spintronic memristive devices”. J.S.R. acknowledges financial support from MICIU/AEI/10.13039/501100011033 through the Severo Ochoa Programme for Centres of Excellence in R\&D (CEX2023-001263-S). J.S. thanks H2020 Marie Skłodowska-Curie grant agreement No. 713673.

\vspace{5mm}

\noindent \textbf{Author contributions}

\noindent J.F.S. and J.S. fabricated the devices with the participation of F.H. using PdSe$_2$ crystals grown by V.M. and D.D.. The measurements were performed by J.F.S, J.S. and W.S.T.. The experimental set-up for angular dependence was enhanced by T.G., who also helped with the measurements. K.X. and J.S.R.  characterized the Raman modes of PdSe$_2$. J.F.S., J.S, L.C. and S.O.V. analyzed the data. All authors commented on the manuscript. S.O.V. conceived and supervised the work, and wrote the manuscript.

\vspace{5mm}

\noindent \textbf{Competing interests}

\noindent The authors declare no competing interests.

\newpage
\noindent \textbf{Figure Captions}
\begin{figure}[ht]
\vspace{10mm}\hspace{10mm}\includegraphics[width=1.05\linewidth]{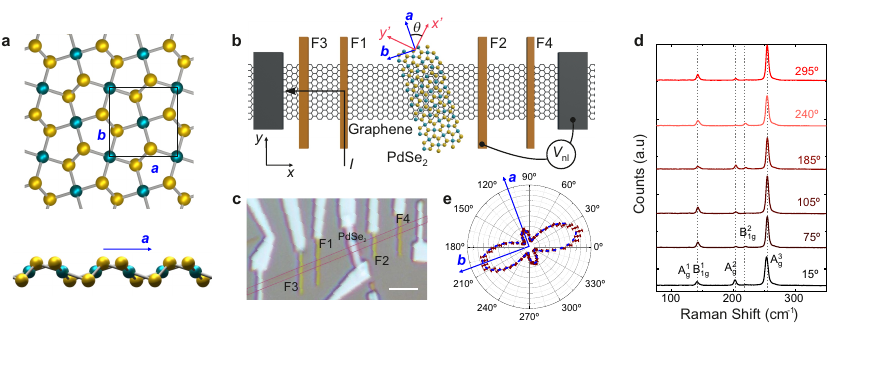}
\vspace{-15mm}
\caption{\setlength{\baselineskip}{0.8\baselineskip}\textbf{Device geometry and monolayer graphene-PdSe$_2$ characterization}. \textbf{a}, Top and side view of the pentagonal lattice of monolayer PdSe$_2$; blue and yellow spheres represent the Pd and Se atoms, respectively. The rectangle denotes the unit cell with lattice constants $a$ and $b$. The side view shows the buckled structure along the $a$-axis. \textbf{b}, Device schematic: monolayer graphene spin channel with four ferromagnetic contacts along $\hat{y}$ (F1-F4, dark orange) and two normal metal contacts (gray). The nonlocal device defined by F1 and F2 probes the proximity effect generated by PdSe$_2$ on graphene. A current $I$ (black arrow) through F1 injects spins into graphene parallel to the F1 magnetization. Spins precess under an external magnetic field while diffusing to the detector F2. Two reference devices (F1-F3 and F2-F4) characterize the pristine graphene. Blue arrows indicate PdSe$_2$ crystalline axes $a$ and $b$; red arrows denote in-plane spin directions $\hat{x}'$ and $\hat{y}'$ for the longest and shortest spin lifetimes, respectively. The angle $\theta$ characterises the rotation between $\hat{x}'$ and $a$. \textbf{c}, Optical image of a graphene-PdSe$_2$ device (Device 1) and the two reference graphene devices. The monolayer graphene flake is highlighted with dotted lines. Scale bar: 5 $\mu$m. \textbf{d}, Representative polarized Raman spectra of the PdSe$_2$ flake in Device 1, showing identified A$_\mathrm{g}$ and B$_{\mathrm{1g}}$ modes. The polarization angle is measured from the $\hat{x}$ axis. The spectra are displaced vertically for clarity. \textbf{e}, Angular dependence of the A$_\mathrm{g}^2$ peak intensity in \textbf{d} to identify the PdSe$_2$ crystalline axes; the error bars represent the standard deviation of the mean value.}
\label{Fig1}
\end{figure}
\newpage
\begin{figure*}[ht!]
\vspace{-10mm}\hspace{5mm}\includegraphics[width=1\linewidth]{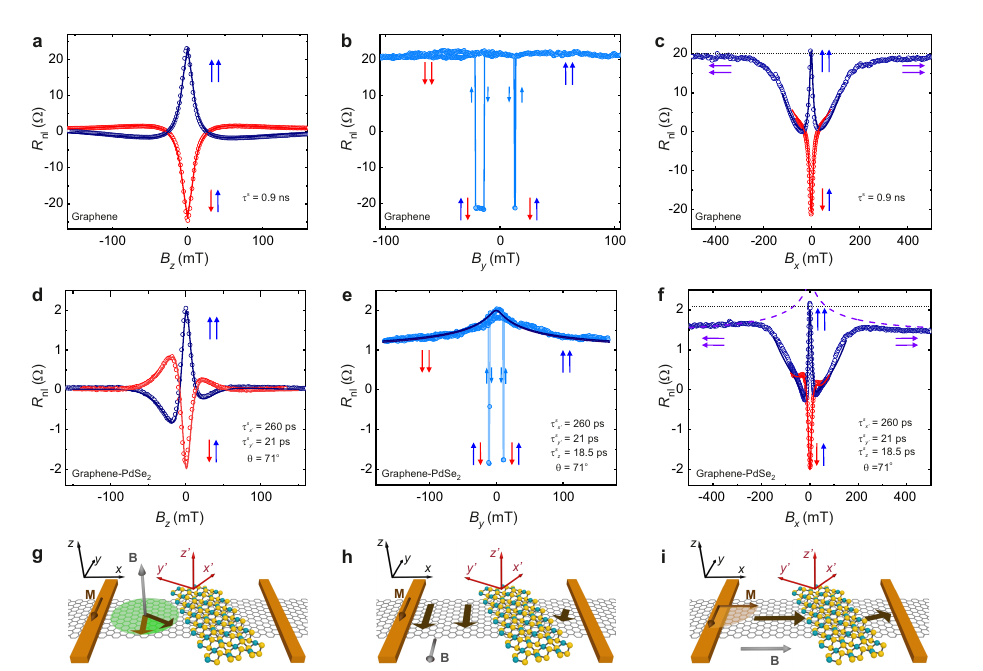}
\vspace{-5mm}
\caption{\setlength{\baselineskip}{0.8\baselineskip}\textbf{Spin relaxation anisotropy in monolayer graphene-PdSe$_2$}. Nonlocal spin resistance $R_\mathrm{nl}\equiv V_\mathrm{nl}/I$ as a function of magnetic field $\mathbf{B}$ along $\hat{x}$, $\hat{y}$ and $\hat{z}$. Measurements in \textbf{a}-\textbf{c} are for the reference device defined by F1 and F3 in Fig. 1c, while those in \textbf{d}-\textbf{f} are for graphene-PdSe$_2$ (Device 1). Open circles represent experimental data, and lines are solutions to the Bloch diffusion equations. Coloured arrows depict the configuration of the ferromagnets' magnetizations. \textbf{a}, \textbf{d}, Spin precession measurements as a function of $B_z$, for parallel (blue) and antiparallel (red) ferromagnet configurations. As illustrated in \textbf{g}, the spins precess in the graphene plane $x-y$. The orientation of the spins at the PdSe$_2$ location is modulated by $B_z$, resulting in asymmetric $R_\mathrm{nl}$ (\textbf{d}), from which $\tau_{x'}^{\mathrm{s}}$, $\tau_{y'}^{\mathrm{s}}$ and $\theta$ can be determined. The asymmetry is absent in isotropic graphene (\textbf{a}).  \textbf{b}, \textbf{e}, $R_\mathrm{nl}$ as a function of $B_y$. As illustrated in \textbf{h}, injected spins, initially parallel to $\hat{y}$ and $\mathbf{B}$, effectively rotate towards $\hat{x}'$ when diffusing under PdSe$_2$. As they misalign from $\mathbf{B}$, they precess out of plane, making $R_\mathrm{nl}$ field-dependent and sensitive to $\tau_{x'}^{\mathrm{s}}$, $\tau_{y'}^{\mathrm{s}}$ and $\tau_{z}^{\mathrm{s}}$ (\textbf{e}). In isotropic graphene, $R_\mathrm{nl}$ is field independent (\textbf{b}). \textbf{c}, \textbf{f}, $R_\mathrm{nl}$ as a function of $B_x$ for parallel (blue) and antiparallel (red) ferromagnet configurations. For low $B_x$, injected spins precess perpendicular to the substrate. As illustrated in \textbf{i}, increasing $B_x$ rotates the ferromagnets magnetization $\mathbf{M}$, and aligns it with $\hat{x}$ for $B_x > B^\mathrm{sat}_x$. In graphene-PdSe$_2$ (\textbf{f}), this causes asymmetric $R_\mathrm{nl}$ about $B_x=0$, similar to \textbf{d}, and a field-dependent reduced  $R_\mathrm{nl}$ for $B_x > B^\mathrm{sat}_x$, as in \textbf{e}.
Solutions of the Bloch equations in \textbf{d}-\textbf{f} assume $\tau_{x'}^{\mathrm{s}}= 260$ ps, $\tau_{y'}^{\mathrm{s}}=21$ ps, $\tau_{z'}^{\mathrm{s}}=18.5$ ps and $\theta=71^\circ$. In \textbf{f}, the dashed line represents the solution for $\mathbf{M}$ aligned with $\hat{x}$, valid for $B_x > B^\mathrm{sat}_x$; solid lines use the Stoner-Wohlfarth approximation with $B^\mathrm{sat}_x = 0.18$ T. Measurements were performed at room temperature with $V_\mathrm{g}=-20$ V (\textbf{a}) and $V_\mathrm{g}=-30$ V (\textbf{b}-\textbf{f}).} \label{Fig2}
\end{figure*}
\newpage
\begin{figure*}[ht!]
\vspace{10mm}\hspace{10mm}\includegraphics[width=1\linewidth]{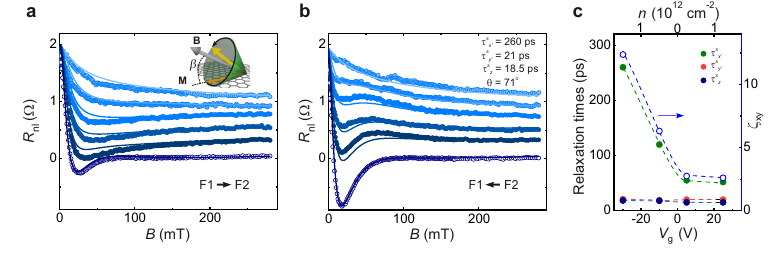}
\caption{\setlength{\baselineskip}{0.8\baselineskip}\textbf{Oblique spin precession and spin lifetime anisotropy ratio $\zeta_{xy}$}. \textbf{a}, Out-of-plane spin precession with F1 as spin injector and F2 as detector for selected $\beta$. Inset: $\mathbf{B}$ is in a plane perpendicular to the substrate containing the ferromagnets' easy axis ($\hat{y}$), its direction is characterised by the angle $\beta$. \textbf{b}, Out-of-plane spin precession with F2 as spin injector and F1 as detector for selected $\beta$. Measurements in \textbf{a} and \textbf{b} are at room temperature for Device 1, with $V_\mathrm{g}=-30$ V. Solid lines are solutions of the Bloch equations with $\tau_{x'}^{\mathrm{s}}= 260$ ps, $\tau_{y'}^{\mathrm{s}}=21$ ps, $\tau_{z'}^{\mathrm{s}}=18.5$ ps and $\theta=71^\circ$ for $\beta = 3^\circ, 20^\circ, 29^\circ, 41^\circ, 51^\circ$ and $90^\circ$ (top to bottom). \textbf{c}, $\tau_{x'}^{\mathrm{s}}$, $\tau_{y'}^{\mathrm{s}}$, $\tau_{z}^{\mathrm{s}}$ and $\zeta_{xy}\equiv\tau_{x'}^{\mathrm{s}}/\tau_{y'}^{\mathrm{s}}$ as a function of $V_\mathrm{g}$. $\tau_{y'}^{\mathrm{s}}$, $\tau_{z}^{\mathrm{s}}$ are largely independent of $V_\mathrm{g}$ and around 20 ps, while $\tau_{x'}^{\mathrm{s}}$ decreases from 260 ps at $V_\mathrm{g}=-30$ V to 52 ps at $V_\mathrm{g}=25$ V, resulting in $\zeta_{xy}$ decreasing from 12.5 to 2.5. The top axis represents the gate-controlled carrier density $n$, with $n=0$ the location of the charge neutrality point of the pristine graphene (see Supplementary Fig. 1). Lines are guides for the eye.
} \label{Fig3}
\end{figure*}
\newpage
\begin{figure*}[ht!]
\vspace{10mm}\hspace{-2mm}\includegraphics[width=1.02\linewidth]{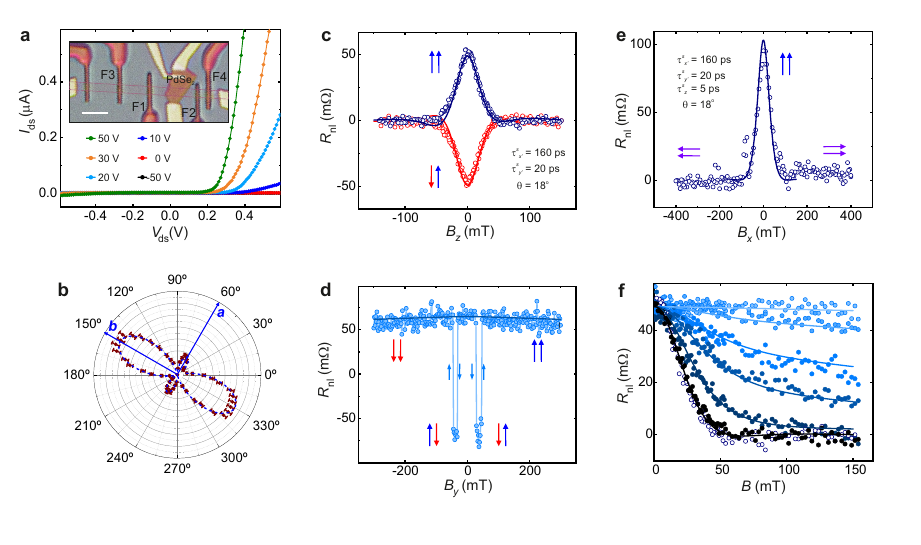}
\vspace{-15mm}
\caption{\setlength{\baselineskip}{0.8\baselineskip}\textbf{Transfer characteristics and spin relaxation anisotropy in monolayer graphene-PdSe$_2$ at low temperature}. \textbf{a}, Transfer characteristics $I_{ds}$ versus bias voltage $V_\mathrm{ds}$ for different $V_\mathrm{g}$ in graphene–PdSe$_2$. $I_{ds}$ is observed only for $V_\mathrm{ds} >0.2$ V. For $V_\mathrm{g}\leq0$, $I_{ds}=0$ over the measured $V_\mathrm{ds}$ range. Inset: Optical image of a graphene-PdSe$_2$ device (Device 2) and the two reference graphene devices. The monolayer graphene flake is highlighted with dotted lines. Scale bar: 5 $\mu$m. \textbf{b}, Angular dependence of the Raman A$_\mathrm{g}^2$ mode intensity to identify the crystalline axes of the PdSe$_2$ flake; the error bars represent the standard deviation of the mean value. \textbf{c}-\textbf{e}, Nonlocal spin resistance $R_\mathrm{nl}$ as a function of magnetic field $\mathbf{B}$ along $\hat{z}$, $\hat{x}$ and $\hat{y}$, respectively. \textbf{f}, Oblique spin precession for $\beta = 0^\circ, 10^\circ, 20^\circ, 30^\circ, 45^\circ, 70^\circ$ (solid circles, top to bottom) and $90^\circ$ (open circles). Measurements in \textbf{a} and \textbf{c}-\textbf{f} were performed at 77 K. For \textbf{c}-\textbf{f}, $V_\mathrm{g}=50$ V. In \textbf{c}, \textbf{d} and \textbf{f} measurements were taken using F1-F4 and in \textbf{e}, using F1-F2. A spin-independent background was subtracted. Solid lines are solutions of the Bloch diffusion equations with $\tau_{x'}^{\mathrm{s}}= 160$ ps, $\tau_{y'}^{\mathrm{s}}=20$ ps, $\tau_{z}^{\mathrm{s}}=5$ ps, $\theta=18^\circ$ and the corresponding $\beta$.
} \label{Fig4}
\end{figure*}

\newpage

\section{Methods}

\noindent \small{\textbf{Device Fabrication}

\noindent PdSe$_2$ single crystals were grown by the self-flux (high-temperature solution) method, resulting in layered crystals with dimensions of 0.8-1 cm$^3$. Excess selenium (Se) was used as a solvent/flux. A preliminary sintering process was carried out in a furnace, with the temperature increased at a rate of  20 $^{\circ}$C per hour from room temperature and 850 $^{\circ}$C, then maintained at 850 $^{\circ}$C for 4 days. To grow the PdSe$_2$ crystals, the ground and homogenized sintered material was placed in a quartz ampoule. The ampoule containing the mixed material was evacuated to a vacuum of 10$^{-5}$ torr and sealed.  The temperature was then raised again to 850 $^{\circ}$C and maintained for 3 days, after which it was decreased at a rate of 1 $^{\circ}$C per hour to 400 $^{\circ}$C. From 400 $^{\circ}$C down to room temperature, the cooling rate was 40 $^{\circ}$C/hour.
Graphene-PdSe$_2$ heterostructures were fabricated by dry viscoelastic stamping [51, 52] inside a glovebox with argon atmosphere. The transfer set-up comprises of an optical microscope with large working distance optical objectives (Nikon LE Plan EPI) and a remote-controlled three-axis micromanipulator. Graphene was obtained by mechanically exfoliating highly oriented pyrolytic graphite (SPI Supplies) onto a $p$-doped Si/SiO$_2$ substrate. Using optical contrast, previously calibrated with Raman spectroscopy, monolayer graphene channels were identified. To fabricate the van der Waals heterostructures, PdSe$_2$ flakes were picked up with the viscoelastic stamp (Gelpack) and transferred onto the graphene, as shown in Fig. 1c and inset of Fig. 2a. After assembling, the stacks were annealed for 3 hours at 290$^{\circ}$C in high vacuum (10$^{-8}$ Torr). The contact electrodes were defined in two subsequent electron-beam lithography steps using resin masks fabricated with spin-coated MMA and PMMA, first for the spin-insensitive electrodes, Ti(1nm)-Pd(30nm), and then for the ferromagnets, TiO$_\mathrm x$/Co(30nm). The contact materials were deposited by electron-beam evaporation in a chamber with a base pressure of 10$^{-8}$ Torr. The TiO$_\mathrm{x}$ barriers were fabricated by evaporating 4 {\AA} + 4 {\AA} of Ti and 30 min oxidation after each evaporation in an oxygen atmosphere of about 10$^{-1}$ Torr.}

\vspace{5mm}

\noindent\small{\textbf{Raman characterization}

\noindent After completing the spin transport characterization, the monolayer nature of the device's graphene channel was confirmed through Raman spectroscopy. The 2D and G bands offer a straightforward method to distinguish between monolayer and multilayer graphene [53]. The 2D band serves as a key indicator of the number of graphene layers. For single layer graphene, the 2D band displays a single symmetric Lorentzian peak with a full width at half maximum of ~30 cm-1. In contrast, bilayer and multilayer graphene exhibit a 2D band that splits into multiple overlapping modes, requiring a fit with several Lorentzian components. Another hallmark of monolayer graphene is the relatively high intensity of the 2D band compared to the G band. Supplementary Fig. 2 presents the Raman spectra of the graphene in the two devices, confirming that the spin channels are indeed composed of monolayer graphene.

\vspace{5mm}

\noindent\small{\textbf{Electrical Characterization}

\noindent The devices are wired to a chip carrier and placed in cryo-free or N$_2$-flow cryostats. Charge transport properties were characterized by employing two- and four-terminal measurements. The contact resistance in the TiO$_\mathrm x$/Co electrodes is about a few 10s of k$\Omega$. The typical average electron/hole mobility in the devices is about $\mu=5000$ cm$^2$V$^{-1}$s$^{-1}$ with a residual carrier density of $3.2\times 10^{11}$ cm$^{-2}$. The carrier density $n$ in the device is controlled with a back-gate voltage applied to the $p$-doped Si substrate.}

\vspace{5mm}

\noindent\textbf{Data Availability} The data that support the plots within the paper and other findings of this study are available online at LINK. Source data are provided with this paper. 

\vspace{5mm}

\noindent\textbf{Code Availability} Source data in the paper are available online at LINK. Any further data and codes used in this study are available from corresponding authors upon reasonable request. 

\vspace{5mm}

\noindent \textbf{Corresponding authors}

\noindent Correspondence to Juan F. Sierra or Josef Sv\v{e}tlík or Sergio O. Valenzuela.


\newpage

\begin{thebibliography}{}

\item[] \noindent \textbf{References}

\vspace{5mm}


\bibitem{RMP2010a} Hasan, M.Z., Kane, C.L. Colloquium: Topological insulators. \textit{Rev. Mod. Phys.} {\bf 82}, 3045 (2010).

\bibitem{RMP2010b} Qi, X.-L., Zhang S.-C. Topological insulators and superconductors. \textit{Rev. Mod. Phys.} {\bf 83}, 1057 (2011).

\bibitem{SOV2015} Sinova, J., Valenzuela, S. O., Wunderlich, J., Back, C. H., \& Jungwirth, T. Spin Hall effects.
\textit{Rev. Mod. Phys.} {\bf 87}, 1213 (2015).

\bibitem{manchon2019} Manchon A., \u{Z}elezn\'{y}, J., Miron, I. M., Jungwirth, T., Sinova, J., Thiaville, A., Garello, K., \& Gambardella, P. Current-induced spin-orbit torques in ferromagnetic and antiferromagnetic systems.
\textit{Rev. Mod. Phys.} {\bf 91}, 035004 (2019).

\bibitem{nadj-perge2010} Nadj-Perge, S., Frolov, S. M., Bakkers, P. A. M., \&  Kouwenhoven, L. P. Spin-orbit qubit in a semiconductor nanowire. \textit{Nature} {\bf 468}, 1084 (2010).

\bibitem{brooks2020} Books, M. \& Burkard, G. Electric dipole spin resonance of two-dimensional semiconductor spin qubits
\textit{ Phys. Rev. B } {\bf 101}, 035204 (2020).

\bibitem{Majoranareview} Prada, E., San-Jose, P., de Moor, M.W.A. \textit{et al.}. From Andreev to Majorana bound states in hybrid superconductor–semiconductor nanowires. \textit{Nat. Rev. Phys.} {\bf 2}, 575-594 (2020).

\bibitem{amundsen2024} Amundsen, M., Linder, J., Robinson, J. W. A., Zutic, I., Banerjee, N. Colloquium: Spin-orbit effects in superconducting hybrid structures. \textit{Rev. Mod. Phys.} {\bf 96}, 021003 (2024).

\bibitem{zutic2020} Zutic, I., Matos-Abiague, A., Scharf, B., Dery, H. \& Belaschhenko, K. Proximitized materials. \textit{Mater. Today } {\bf 22}, 85–107 (2019).

\bibitem{JFS2021} Sierra, J. F., Fabian, J., Kawakami, R. K., Roche, S. \& Valenzuela, S. O. Van der Waals heterostructures for spintronics and opto-spintronics. \textit{Nat. Nanotechnol.} {\bf 16}, 856–868 (2021).

\bibitem{ghiasiMPE} Ghiasi, T.S., Kaverzin, A.A., Dismukes, A.H. \textit{et al.}. Electrical and thermal generation of spin currents by magnetic bilayer graphene. \textit{Nat. Nanotechnol.} {\bf 16}, 788–794 (2021).

\bibitem{zhongMPE} Zhong, D., Seyler, K.L., Linpeng, X. \textit{et al.}. Layer-resolved magnetic proximity effect in van der Waals heterostructures. \textit{Nat. Nanotechnol.} {\bf 15}, 187–191 (2020).

\bibitem{lyons2020} Lyons, T. P. \textit{et al.}. Interplay between spin proximity effect and charge-dependent exciton dynamics in MoSe$_2$/CrBr$_3$ van der Waals heterostructures. \textit{Nat. Commun.} {\bf 11}, 6021 (2020).

\bibitem{wang2020} Wang H. \textit{et al.}. Above room-temperature ferromagnetism in wafer-scale two-dimensional van der Waals Fe$_3$GeTe$_2$ tailored by a topological insulator. \textit{ACS Nano} \textbf{14}, 10045-10053 (2020).

\bibitem{zhangSC2023} Zhang, Y., Polski, R., Thomson, A. \textit{et al.}. Enhanced superconductivity in spin–orbit proximitized bilayer graphene. \textit{Nature} {\bf 613}, 268–273 (2023).

\bibitem{KS2024} Denisov, K.S., Rozhansky, I.V., Valenzuela, S.O., Zutic, I. Terahertz spin-light coupling in proximitized Dirac materials. \textit{Phys. Rev. B} {\bf 109}, L201406 (2024).

\bibitem{gmitra2015} Gmitra, M. \& Fabian, J. Graphene on transition-metal dichalcogenides: A platform for proximity spin-orbit
physics and optospintronics.
\textit{ Phys. Rev. B } {\bf 92}, 155403 (2015).

\bibitem{gmitra2016} Gmitra, M., Kochan, D., H\"{o}gl, P. \& Fabian, J. Trivial and inverted Dirac bands and the emergence of quantum spin Hall states in graphene
on transition-metal dichalcogenides.
\textit{Phys. Rev. B} {\bf 93}, 155104 (2016).

\bibitem{ghiasi2017} Ghiasi, T. S., Ingla-Aynés, J., Kaverzin, A. A., \& van Wees, B. J. Large Proximity-induced spin lifetime anisotropy in transition-metal dichalcogenide/graphene heterostructures
\textit{Nano Lett.} \textbf{17}, 7528-7532 (2017).

\bibitem{LAB2018}  Ben\'{\i}tez, L. A., Sierra, J. F., Savero Torres, W., Arrighi, A., Bonell, F., Costache, M. V. \& Valenzuela, S. O. Strongly anisotropic spin relaxation in graphene-transition metal dichalcogenide heterostructures at room temperature.
\textit{Nat. Phys.} {\bf 14}, 303-308 (2018).

\bibitem{safeer2019} Safeer, C. K., Ingla-Ayn\'{e}s, J., Herling, F., Garcia, J. H., Vila, M., Ontoso, N., Calvo, M. R., Roche, S., Hueso, L. E. \& Casanova, F. Room-temperature spin Hall effect in graphene/MoS$_2$ van der Waals heterostructures
\textit{Nano Lett.} \textbf{19}, 1074-1082 (2019).

\bibitem{ghiasi2019} Ghiasi, T.S., Kaverzin, A. A., Blah, P. J., \& van Wees, B. J.
Charge-to-spin conversion by the Rashba-Edelstein effect in 2D van der Waals heterostructures up to room temperature.
\textit{Nano Lett.} \textbf{19}, 5959-5966 (2019).

\bibitem{LAB2020} Ben\'{\i}tez, L. A., Savero Torres, W., Sierra, J. F. \textit{et al.}. Tunable room-temperature spin galvanic and spin Hall effects in van der Waals heterostructures. \textit{Nat. Mater.} {\bf 19}, 170–175 (2020).

\bibitem{RG2021} Galceran, R., Tian, B., Li, J., \textit{et al.}.  Control of spin–charge conversion in van der Waals heterostructures. \textit{APL Mater.} {\bf 9}, 100901 (2021).

\bibitem{herling2020} Herling, F., Safeer, C. K., Ingla-Ayn\'{e}s, J., Ontoso, N., Hueso, L. E., \& Casanova F. Gate tunability of highly efficient spin-to-charge conversion by spin Hall effect in graphene proximitized with WSe$_2$. \textit{APL Mater.} {\bf 8}, 071103 (2020).

\bibitem{cummings2017} Cummings, A. W., Garcia, J. H., Fabian, J.,\& Roche, S. Giant Spin Lifetime Anisotropy in Graphene Induced by Proximity Effects.
\textit{Phys. Rev. Lett.} {\bf 119}, 206601 (2017).

\bibitem{offidani2018} Offidani, M. \& Ferreira, A. Microscopic theory of spin relaxation anisotropy in graphene with proximity-induced spin–orbit coupling.
\textit{Phys. Rev. B} {\bf 98}, 245408 (2018).

\bibitem{safeer2019b} Safeer, C. K., Ontoso, N., Ingla-Ayn\'{e}s, J., Herling, F., \textit{et. al}. Large Multidirectional Spin-to-Charge Conversion in Low-Symmetry Semimetal MoTe$_2$ at Room Temperature
\textit{Nano Lett.} \textbf{19}, 8758-8766 (2019).

\bibitem{hoque2021} Hoque, A. Md., Khokhriakov, D., Zollner, K., Zhao, B., Karpiak, K.,  Fabian, J. \& Dash, S. P. All-electrical creation and control of spin-galvanic signal in graphene and molybdenum ditelluride heterostructures at room temperature.
\textit{Commun. Phys.} \textbf{4}, 87124 (2021).

\bibitem{zhao2020} Zhao, B., Karpiak, B., Khokhriakov, D., Johansson, A., Hoque, A. Md., Xu, X., Jiang, Y., Mertig, I. \& Dash, S. P. Unconventional charge–spin conversion in Weyl-semimetal WTe$_2$
\textit{Adv. Mat.} \textbf{32}, 2000818 (2020).

\bibitem{LC2022} Camosi, L., Světlík, J., Costache, M. V., Savero Torres, W., Fernández Aguirre, I., Marinova, V., Dimitrov, D., Gospodinov, M., Sierra, J. F., Valenzuela, S. O. Resolving spin currents and spin densities generated by charge-spin interconversion in systems with reduced crystal symmetry.
\textit{2D Mat.} \textbf{9}, 035014 (2022).

\bibitem{cording2024} Cording, L., Liu, J., Tan, J. Y. \textit{et al.}. Highly anisotropic spin transport in ultrathin black phosphorus.
\textit{Nature} {\bf 23}, 479-485 (2024).

\bibitem{newfabian} Milivojević, M., Gmitra, m., Kurpas, M., Štich, I., Fabian, J. Giant asymmetric proximity-induced spin-orbit coupling in twisted graphene/SnTe heterostructure. \textit{2D Matter.} {\bf 11}, 035036 (2024).

\bibitem{sun2015} Sun, J., Shi, H., Siegrist, T., Singh, D. J. Electronic, transport, and optical properties of bulk and mono-layer PdSe$_2$. \textit{Appl. Phys. Lett.} {\bf 107}, 153902 (2015).

\bibitem{oyedele2017} Oyedele, A.D., Yang. S., Liang, L., \textit{et al.}. PdSe$_2$: Pentagonal Two-Dimensional Layers with High Air Stability for Electronics. \textit{J. Am. Chem. Soc.} {\bf 139}, 14090-14097 (2017).

\bibitem{david2019} David, A., Rakyta, P., Kormanyos, A. \& Burkard, G. Induced spin-orbit coupling in twisted graphene–transition metal dichalcogenide heterobilayers: Twistronics meets spintronics.
\textit{Phys. Rev. B} {\bf 100}, 085412 (2019).

\bibitem{BR2016} Raes, B., Scheerder, J. E., Costache, M. V., Bonell, F., Sierra, J. F., Cuppens, J., Van de Vondel, J., \& Valenzuela S. O. Determination of the spin-lifetime anisotropy in graphene using oblique spin precession.
\textit{Nat. Commun.} {\bf 7}, 11444 (2016).

\bibitem{BR2017} Raes, B., Cummings, A., Bonell, F., Costache, M. V., Sierra, J. F., Roche, S., \& Valenzuela S. O. Spin precession in anisotropic media.
\textit{Phys. Rev. B} {\bf 95}, 085403 (2017).

\bibitem{LAB2019} Ben\'{\i}tez, L. A., Sierra, J. F., Savero Torres, W., Timmermans, M., Costache, M. V. \& Valenzuela, S. O. Investigating the spin-orbit interaction in van der Waals heterostructures by means of the spin relaxation anisotropy. \textit{APL Mater.} {\bf 7}, 120701 (2019).

\bibitem{yu2020} Yu, J., Kuang, X., Gao, Y., \textit{et al.}. Direct Observation of the Linear Dichroism Transition in Two-Dimensional Palladium Diselenide. \textit{Nano Lett.} {\bf 20}, 1172-1182 (2020).

\bibitem{yan2016} Yan, W. \textit{et al.}. A two-dimensional spin field-effect switch. \textit{Nat. Commun.} {\bf 7}, 13372 (2016).

\bibitem{dankert2017} Dankert, A. \& Dash, S. Electrical gate control of spin current in van der Waals heterostructures at room temperature. \textit{Nat. Commun.} {\bf 8}, 1609 (2017).

\bibitem{naimer2021} Naimer, T., Zolner, K, Gmitra, M., Fabian, J. Twist-angle dependent proximity induced spin-orbit coupling in graphene/transition metal dichalcogenide heterostructures. \textit{Phys. Rev. B} {\bf 104}, 195156 (2021).

\bibitem{MV2021} Vila, M., Hsu, C. H., Garcia, J. H., \textit{et al.}. Low-symmetry topological materials for large charge-to-spin interconversion: The case of transition metal dichalcogenide monolayers. \textit{Phys. Rev. Research} {\bf 3}, 043230 (2021).

\bibitem{kurpas2018} Kurpas, M., Gmitra, M., \& Fabian, J. Spin properties of black phosphorus and phosphorene, and their prospects for spincalorics. \textit{J. Phys. D: Appl. Phys.}  {\bf 51}, 17401 (2018).

\bibitem{morpurgo2015} Wang, Z., Ki, D., Chen, H. \textit{et al.}. Strong interface-induced spin–orbit interaction in graphene on WS2. \textit{Nat. Commun.} {\bf 6}, 8339 (2015).

\bibitem{morpurgo2016} Wang, Z., Ki, D. -K., Khoo, J. Y. \textit{et al.}. Origin and magnitude of ‘designer’ spin-orbit interaction in graphene on semiconducting transition metal dichalcogenides. \textit{Phys. Rev. X} {\bf 6}, 041020 (2016).

\bibitem{bouchiat2018} Wakamura, T., Reale, F., Palczynski, P. \textit{et al.}. Strong anisotropic spin-orbit interaction induced in graphene by monolayer WS$_2$. \textit{Phys. Rev. Lett.} {\bf 120}, 106802 (2018).

\bibitem{wang2019} Wang, D. \textit{et al.}. Quantum Hall effect measurement of spin-orbit coupling strengths in ultraclean bilayer graphene/WSe$_2$ heterostructures. \textit{Nano Lett.} {\bf 19}, 7028  (2019).

\bibitem{monaco2021} Monaco, C., Ferreira, A., Raimondi, R. Spin Hall and inverse spin galvanic effects in graphene with strong interfacial spin-orbit coupling: A quasi-classical Green's function approach. \textit{Phys. Rev. Research} {\bf 3}, 033137  (2021).

\end{thebibliography}

\vspace{5mm}


\begin{thebibliography}{}

\item[] \noindent \textbf{Methods-only references}

\vspace{5mm}

\item[] [51] Castellanos-Gomez, A. \textit{et al.}. Deterministic transfer of two-dimensional materials by all-dry viscoelastic stamping. \textit{2D Mater.} {\bf 1}, 011002 (2014).

\item[] [52] Frisenda, R., Navarro-Moratalla, E., Gant, P. \textit{et al.}. Recent progress in the assembly of nanodevices and van der Waals heterostructures by deterministic placement of 2D materials. \textit{Chem. Soc. Rev.} {\bf 47}, 53-68 (2018).

\item[] [53] Malard, L. M., Pimenta, M. A., Dresselhaus, G., \& Dresselhaus, M. S. \textit{Phys. Rep.} {\bf 473}, 51-87 (2009).

\end{thebibliography}
\end{document}